\documentclass[aps,prd,onecolumn,superscriptaddress]{revtex4}
\usepackage{graphicx}
\usepackage{amsmath}
\usepackage{epsfig, epstopdf}
\usepackage{amsmath,amsfonts,amssymb}
\usepackage[export]{adjustbox}
\usepackage{blindtext}
\usepackage{epsf}

\usepackage{color}

\begin{document}
\baselineskip=12pt
\def\black{\textcolor{black}}
\def\red{\textcolor{black}}
\def\blue{\textcolor{blue}}
\def\green{\textcolor{black}}
\def\be{\begin{equation}}
\def\ee{\end{equation}}
\def\bea{\begin{eqnarray}}
\def\eea{\end{eqnarray}}
\def\orc{\Omega_{r_c}}
\def\om{\Omega_{\text{m}}}
\def\E{{\rm e}}
\def\bearst{\begin{eqnarray*}}
\def\eearst{\end{eqnarray*}}
\def\peleven{\parbox{11cm}}
\def\peffec{\peight{\bearst\eearst}\hfill\peleven}
\def\pspace{\peight{\bearst\eearst}\hfill}
\def\ptwelve{\parbox{12cm}}
\def\peight{\parbox{8mm}}

\title{Lensing as a Probe of Early Universe: from CMB to Galaxies }

\author{Farbod Hassani}
\email{farbod-AT-physics.sharif.edu}
\address{Department of Physics, Sharif University of
Technology, P.~O.~Box 11155-9161, Tehran, Iran}

\author{Shant Baghram}
\email{baghram-AT-sharif.edu}
\address{Department of Physics, Sharif University of
Technology, P.~O.~Box 11155-9161, Tehran, Iran}
\author{Hassan Firouzjahi}
\email{firouz-AT-ipm.ir}
\address{School of Astronomy, Institute for Research in
Fundamental Sciences (IPM), P.~O.~Box 19395-5531,Tehran, Iran}

\vskip 1cm

\begin{abstract}
The Cosmic Microwave Background (CMB) radiation lensing is a promising tool to study the physics of early universe. In this work we probe the imprints  of  deviations from isotropy and scale invariance of primordial curvature perturbation power spectrum  on CMB lensing potential and convergence. Specifically, we consider a scale-dependent hemispherical asymmetry in  primordial power spectrum.  We show that the CMB lensing potential and convergence and also the
cross-correlation of the CMB lensing and late time galaxy convergence can probe the amplitude and the scale dependence of the dipole modulation. As another example, we consider a primordial power spectrum with local feature. We show that the CMB lensing and the cross-correlation of the CMB lensing and galaxy lensing can probe the amplitude and the shape of the local feature. We show that the cross correlation of CMB lensing convergence and galaxy lensing  is capable to probe the effects of local features in power spectrum on smaller scales than the CMB lensing. Finally we showed that the current data can constrain the amplitude and moment dependence of dipole asymmetry.

\end{abstract}

\maketitle


\section{INTRODUCTION}

The standard model of Cosmology known as $\Lambda$CDM is established as a cornerstone of modern cosmology. The Cosmic Microwave Background  (CMB) radiation \cite{Planck:2015xua} and Large Scale Structure (LSS) surveys \cite{Tegmark:2003ud,Eisenstein:2005su,Tegmark:2006az} indicate that almost all observations can be described by a 6-parameter $\Lambda$CDM model with an initial condition of almost Gaussian, isotropic, scale invariant and adiabatic perturbations which can be sourced during inflation  \cite{Guth:1980zm, Linde:1981mu, Albrecht:1982wi, Sato:1980yn, Starobinsky:1980te}. Recent observations of Planck collaboration shows almost no deviation from standard picture of single field slow roll inflation \cite{Ade:2015lrj}. On the other hand the BICEP2/Keck - Planck results shows that there is no detection of gravitational waves \cite{Ade:2015tva,Array:2015xqh}. Despite the success of the standard model, the physics of early universe, the natures of dark matter and dark energy are still unknown.

The large scale structure surveys open a new horizon to test the cosmological models in different redshifts and sub-CMB scales. One of the recent developments of the field is the detection of the lensed CMB map with unprecedented accuracy \cite{Keisler:2011aw,Das:2013zf,Ade:2013tyw,Ade:2015zua}. The CMB lensing is caused by the gravitational effect of structures in the line of sight of the photons from the last scattering surface \cite{Lewis:2006fu}. Each ray of  CMB is deflected due to gravitational lensing effect and accordingly we observe the temperature of each point on the  lensed CMB Map,  $\tilde{T}(\hat{n})$,  as a map of the temperature in a nearby point $T(\hat{n}')$ (note that in this work a quantity accompanied with  the symbol $\, \tilde{}\, $ represents the corresponding lensed quantity). The lensing potential  encapsulates all information on the deflection of light bundle  due to  lensing and it is related to the integrated potential of the matter distribution in all redshift up to the surface of  last scattering.

The calculation of the two point statistics in lensed CMB maps can be a probe of the matter distribution (matter power spectrum) in the Universe. Accordingly, the CMB lensing is a unique observational tool to constrain the cosmological models, specially the evolution of the growth of the structures \cite{Ade:2015rim}.
The CMB lensing can not only change the temperature map of CMB but also it can change the polarization maps by being a new source of B-mode polarization \cite{Lewis:2006fu}. In addition, it introduces  non-Gaussianities \cite{Hanson:2009kg}. Therefore, in order to study the primordial Universe, we should  subtract the effects of CMB lensing on CMB map correctly in order to avoid the degeneracies \cite{Amendola:2014wma}. Having said that,  the CMB lensing in its own can be used as a unique observational tool to constrain the physics of early universe \cite{Okamoto:2003zw,Pearson:2012ba}. It can probe the curvature perturbations power spectrum and capture any deviation from standard assumptions.

In this work we use the CMB lensing as a probe of early Universe physics, specially to study the effects of deviation from scale invariant and isotropic initial conditions. This is motivated from the fact that  there are indications of anomalies in CMB data (like power deficit, cold spot, alignment of quadrupole and octupole \cite{Ade:2013nlj}).  In addition, the Planck data indicates the existence of  hemispherical asymmetry in  primordial power spectrum \cite{Ade:2015hxq}, first seen in WMAP data \cite{Eriksen:2003db, Eriksen:2006xr, Eriksen:2007pc},  and also verified with the local variance estimator \cite{Akrami:2014eta}. If these anomalies are not the statistical flukes, then it is worth to investigate their effects on CMB- and
LSS-related observations in order to understand their origins. We propose that the CMB lensing and also CMB lensing cross-correlation with cosmic shear can be used as a unique probe for detection of these anomalies and their scale dependence.

The structure of the work is as follows. In Sec. \ref{Sec:Theory} we review the theoretical background, which is needed for CMB lensing, cosmic shear and their cross-correlation. In Sec. \ref{Sec:Deviation} we study the effects of deviation from isotropy, specifically the dipole modulation in power spectrum, and the deviation from simple scale-invariance via a localized feature in power spectrum. The conclusions and discussions are presented in Sec. \ref{Sec-Conclusion}. Also in App.(\ref{app1}) and App. (\ref{app2}) we probe the parameter space of dipole modulation and Gaussian feature. In this work we use the cosmological parameters of the base $\Lambda CDM$ model from Planck 2015 \cite{Planck:2015xua} results:
$\Omega_bh^2=0.022$, $\Omega_ch^2=0.1198$, $\Omega_{\Lambda}=0.6844$, $n_s=0.96$, $A_s=2.20\times 10^{-9}$ and $H_0=67.27 km/s/Mpc$.


\section{Theoretical Background}
\label{Sec:Theory}
In this section we review the theoretical background for the mechanism of lensing. In first subsection we focus on CMB lensing. The cross-correlation of CMB lensing with cosmic shear is the subject of the study in the second sub-section.

The geometry of the Universe is described by the perturbed FRW metric in "Conformal Newtonian" gauge
\be
ds^2=a^2(\eta)\Big[-\left(1+2\Psi(\vec{x},t) \right)d\eta^2+ \left(1-2\Phi(\vec{x},t) \right)d\chi^2 \Big] \, ,
\ee
where $\eta$ is the conformal time, $\chi$ is the comoving distance and $\Psi$, $\Phi$ are the Bardeen potentials, the
scalar degrees of freedoms for perturbations. We assume that $\Phi=\Psi$, which is the case if the cosmic fluid does not have anisotropic stress and General Relativity (GR) is the correct classical theory of gravity.

\subsection{CMB lensing}

The light source from a distant objects in Universe is distracted by the intervening structures before the line of sight. A crucial parameter that shows the amount of convergence/deconvergence of light bundles is the lensing potential $\phi$ defined as \cite{Lewis:2006fu}:
\be
\phi(\hat{n})=-2\int_0^{\chi_*}d\chi \left(\frac{\chi_*-\chi}{\chi_*\chi} \right)\Psi(\vec{x},\eta ) \, ,
\ee
where the integration parameter $\chi$ is the comoving distance, $\chi_*$ is the comoving distance to the source
and $\hat{n}$ represents the direction of the observed CMB light.

By the knowledge of the lensing potential, we can calculate the deflection angle as $\hat{\alpha}=\nabla_{\hat{n}}\phi(\hat{n})$. The corresponding lensed temperature $\tilde{T}$ is related to the un-lensed temperature $T$ via the effect of  deflection as: $\tilde{T}(\hat{n})=T(\hat{n}')=T(\hat{n}+\hat{\alpha})$. Note that $\nabla_{\hat{n}}$ is the angular gradient, consequently the gravitational potential, the lensing potential and the deflection angle are dimensionless quantities.  Now we can relate the two point function of the lensing potential to the power spectrum of gravitational potential in Fourier space as
\be \label{Eq:lenscor}
\Big\langle\phi(\hat{n})\phi^*(\hat{n}')\Big\rangle=4\int_0^{\chi_*}d\chi\int_0^{\chi_*}d\chi'\int\frac{d^3k}{(2\pi)^3}\int\frac{d^3k'}{(2\pi)^3} \left(\frac{\chi_*-\chi}{\chi_*\chi} \right) \left(\frac{\chi_*-\chi'}{\chi_*\chi'} \right) e^{i\bf{k}.\chi}e^{-i\bf{k}'.\chi'} \Big \langle\Psi({\bf{k}},\eta(\chi))\Psi^*({\bf{k'}},\eta'(\chi')) \Big \rangle \, ,
\ee
where $\Psi(k,\eta)$ is the Fourier transform of gravitational potential.

 We can relate the gravitational potential  $\Psi$ to the primordial curvature perturbation ${\cal{R}}$ via transfer function ${\cal{T}}(k,\eta)$ as:
\be  \label{Eq:transfer}
\Psi({\bf{k}},\eta)={\cal{T}}({\bf{k}},\eta){\cal{R}}({\bf{k}}) \, .
\ee
Correspondingly, the lensing potential will be related to the primordial curvature perturbation $P_{\cal{R}}$ as follows
\be
\Big\langle\phi(\hat{n})\phi^*(\hat{n}')\Big\rangle=4\int_0^{\chi_*}d\chi\int_0^{\chi_*}d\chi'\int\frac{d^3k}{(2\pi)^3}(\frac{\chi_*-\chi}{\chi_*\chi})(\frac{\chi_*-\chi'}{\chi_*\chi'})e^{i\bf{k}.(\chi-\chi')}{\cal{T}}(k,\eta(\chi)){\cal{T}}(k,\eta'(\chi'))P_{\cal{R}}(k) \label{116}
\ee
in which the primordial power spectrum $P_{\cal{R}}$ is defined via
\be \label{Eq:power}
\langle {\cal{R}}({\bf{k}}){\cal{R}}^*({\bf{k}}') \rangle=(2 \pi)^3P_{\cal{R}}(k)\delta^{3}(\bf{k}-\bf{k}') \, .
\ee

Using the definition of the flat wave  and the orthogonality of spherical harmonics, the cross-correlation of lensing potential becomes
\begin{align}
\Big \langle\phi(\hat{n})\phi^*(\hat{n}') \Big\rangle &=4\int_0^{\chi_*}d\chi\int_0^{\chi_*}d\chi'\int\frac{k^2dk}{(2\pi)^3}\left(\frac{\chi_*-\chi}{\chi_*\chi} \right) \left(\frac{\chi_*-\chi'}{\chi_*\chi'} \right){\cal{T}}\left( k,\eta(\chi) \right){\cal{T}}\left( k,\eta'(\chi') \right)\frac{2\pi^2}{k^3}{\cal{P}_{\cal{R}}}(k)  \nonumber \\
 &~~~~~~~~~~~~~~~~~~~~~~~~~~~~~~~~~~~~~~~~~ \times \sum_{lm}(4\pi)^2j_{\l}(k\chi)j_{l}(k\chi')Y_{lm}(\hat{n})Y^*_{lm}(\hat{n}') \label{120} \, ,
\end{align}
where ${\cal{P}}_{\cal{R}} = k^3 P_{\cal R}/2 \pi^2$ is the dimensionless curvature power spectrum with its amplitude and
scale dependence being constrained by  the CMB data \cite{Planck:2015xua}.

The lensing potential can be expanded in terms of spherical harmonics as well
\begin{align}
\phi(\widehat{n})  =\sum_{lm}\phi_{lm}Y_{lm}(\widehat{n}) \label{123} \, .
\end{align}
For a statistically isotropic field, the angular power spectrum is defined as
\begin{align}
\Big\langle \phi_{lm}\phi^*_{l'm'} \Big\rangle &= \delta_{ll'}\delta_{mm'}C_l^{\phi\phi}\label{130}\\
\Big\langle\phi(\hat{n})\phi^*(\hat{n}')\Big\rangle &= \sum_{lm}Y_{lm}(\hat{n})Y^*_{lm}(\hat{n}')C^{\phi\phi}_{l}
\end{align}
where $C_{l}^{\phi\phi}$ is the angular potential of the lensing.

Now considering the fact that the integral over $\chi$ and $\chi'$ is symmetric and interchangeable  from Eq. (\ref{120})
we obtain
\be
C^{\phi\phi}_{l}=16\pi\int \frac{dk}{k}{\cal{P}}_{\cal{R}}(k) \Bigg(\int_0^{\chi_*}d\chi(\frac{\chi_*-\chi}{\chi_*\chi}){\cal{T}}(k,\eta)j_{l}(k\chi) \Bigg)^2  
\label{eq:clpsi} \, .
\ee

Eq.(\ref{eq:clpsi}) shows that the angular power spectrum of CMB lensing depends on the primordial power spectrum ${\cal{P}}_{\cal{R}}(k)$. We emphasis that in obtaining Eq.(\ref{eq:clpsi}) we have assumed that the appropriate transfer function  ${\cal{T}}(k,\eta)$ used in this equation is not modified  compared to the base $\Lambda$CDM universe.
This is justified because we assume that anomalies of our interest, such as dipole asymmetry or local feature,  happen during inflation and
Universe follows its traditional history after inflation.   Accordingly, at least to leading order,
all the effects of the primordial Universe are imprinted in primordial power and they can be investigated in CMB lensing.

In CMB lensing studies we can interchangeably  express the lensing power in terms of the late time matter power spectrum and to test the predictions of the models for matter distribution of late time Universe.
In this light, CMB lensing is a probe of early universe through Eq.(\ref{eq:clpsi}) and also the distribution of the matter in late time.

In below we discuss the theoretical background of the relation between the CMB lensing potential power and the matter power spectrum. Defining the  gravitational potential power spectrum $P_{\Psi}(k,\chi,\chi')$  via
\be \label{Eq:powerpot}
\langle \Psi({\bf{k}},\chi)\Psi^*({\bf{k}}',\chi') \rangle=(2\pi)^3P_{\Psi}(k,\chi,\chi')\delta^{3}({\bf{k}}-{\bf{k}}').
\ee
The angular power spectrum of lensing becomes

\be
C_l^{\phi\phi}=\frac{8}{\pi}\int k^2dk\int_0^{\chi_*}d\chi\int_0^{\chi_*}d\chi P_{\Psi}(k;\chi,\chi')j_{l}(k\chi)j_{l}(k\chi')  \left(\frac{\chi_*-\chi}{\chi_*\chi} \right) \left(\frac{\chi_*-\chi'}{\chi_*\chi'} \right) \, .
\ee
Now using the Poisson equation  we can relate the power of gravitational potential  to the matter power as follows
\be
P_{\Psi}(k,\chi)= \frac{9}{4}\frac{H_0^4\Omega_m^2(1+z)^2}{k^4}P_{\delta}(k,z) \, ,
\ee
where $\Omega_m$ is the density parameter of matter in present time, $P_{\delta}(k,z)$ is the matter power spectrum which can be expressed in terms of growth function $D(z)$ and the matter transfer function $T(k)$ via
\be
P_{m}(k,z) = A T^2(k) D^2(z) k^{n_s} \, ,
\ee
where $n_s$ is the spectral index.  In this work we use the transfer function of Eisenstein and Hu \cite{Eisenstein:1997ik} for transfer function  which captures the baryon acoustic oscillations.

In the high moment  approximation $(l \gg 1)$, we can use the Limber formula $\int k^2 dk j_l(k\chi)j_l(k\chi')=\left( \pi/\chi^2 \right) \delta(\chi-\chi')$ to study the lensing potential in small angular scales, yielding
\be
C_l^{\phi\phi} \simeq 9\int_0^\infty\frac{d\chi}{\chi^2}(1+z)^2\frac{H_0^4\Omega_m^2}{k^4}(\frac{\chi_*-\chi}{\chi_*\chi})^2P_m(k,z) \, .
\ee
Now using the fact that in small angle limit the wavenumber becomes $k= l/\chi$ and also by exchanging the variable of
integral from comoving distance to redshift results in
\be
l^4C_l^{\phi\phi}=\int_0^{z_*}dzW^{\phi}(z)P_m\Big(\frac{l}{\chi(z)},z \Big),
\ee
where $W^{\phi}$ is the kernel of lensing potential obtained as
\be \label{eq:Wphi}
W^{\phi}(z)=9\frac{(cH_0^{-1})^{-3}}{E(z)}\chi^2 (\frac{\chi_*-\chi}{\chi\chi_*})^2\Omega_m^2(1+z)^2 \, ,
\ee
in which $E(z)\equiv H(z)/H_0$.
We can also find the angular power spectrum of convergence $\hat{k}$, which is defined as follows
\be
\hat{\kappa} \equiv \frac{1}{2}\nabla^2_{\hat{n}}\phi(\hat{n}) \, .
\ee
Accordingly,  the angular power spectrum $C_l^{\kappa_c\kappa_c}$ is obtained to be
\be \label{Eq:powerKcmb}
C_l^{\kappa_c\kappa_c}=\frac{9}{4}\int_0^{z_*}\frac{dz}{E(z)}(1+z)^2H_0^3\Omega_m^2\chi^2 \left(\frac{\chi_*-\chi}{\chi_*\chi} \right)^2P_m(k,z) \, .
\ee
Worth to mention that observationally both the lensing and convergence power spectrum are used to constrain the cosmological models. In the next sections we study the deviation from isotropic and scale invariant initial conditions
for primordial power spectrum.

\subsection{Cosmic shear and cross correlation with CMB lensing}


\begin{figure}
\advance\leftskip-1.5cm
\includegraphics[scale=1.2]{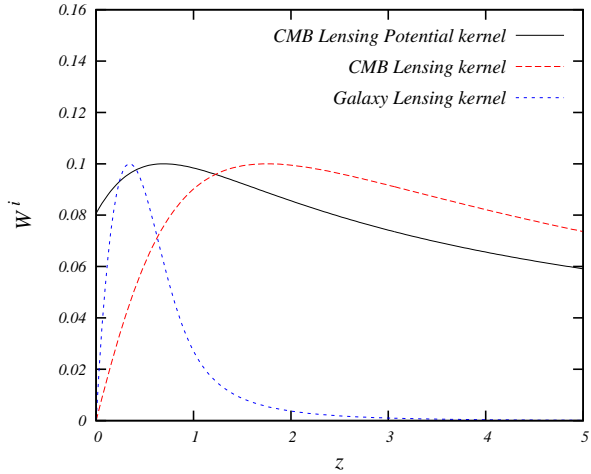}
\caption{ The lensing kernels $W^{i}$ versus redshift are plotted  where $i$ represent each of the kernels. The black solid line represents the CMB lensing potential kernel ($i=\phi$) introduced in Eq.(\ref{eq:Wphi}). The red long dashed line shows the CMB lensing convergence kernel ($i=\kappa_{c}$) introduce in Eq.(\ref{eq:Wkcmb}). The blue dashed line shows the galaxy lensing convergence kernel ($i=\kappa_{g}$) introduced in Eq.(\ref{eq:Wkg}). For $W^{\kappa_g}$ we used the galaxy distribution from Hand et al. \cite{Hand:2013xua}.}
\label{Kernels}
\end{figure}

On cosmological scales besides the CMB lensing there is another lensing effect known as the cosmic shear. This effect is the distortion of the light of the source galaxies in weak gravitational lensing limit by the  surrounding matter distribution \cite{Bartelmann:1999yn,Bernardeau:1999mh,Refregier:2003ct}. This effect has been seen in different weak lensing surveys \cite{Bacon:2000sy,Wittman:2000tc,Fu:2007qq}. In order to study the weak lensing effect of structures in late time, we define the convergence as the weighted integral of matter density contrast in line of sight as follows
\be
\kappa(\hat{n})=\int_0^\infty dz W(z)\delta \left(\chi(z)\hat{n},z \right) \, .
\ee
Here $W(z)$ is the weighting function which is related to the cosmological model and to the  distribution of the sources via
\be \label{eq:Wkg}
W^{\kappa_g}(z)=\frac{3}{2}\Omega_m H_0^2\frac{1+z}{H(z)}\chi(z)\int_z^{\infty}dz\frac{dn(z)}{dz_s}\frac{\chi(z_s)-\chi(z)}{\chi(z_s)} \, ,
\ee
where  $z_s$ represents the redshift of the source and $dn(z)/dz_s$ is the distribution of sources in redshift space. In the case of the CMB lensing, the source is localized at $z_*$ and we obtain Eq. (\ref{Eq:powerKcmb}) with the kernel of CMB lensing $W^{\kappa_{c}}$ defined as below
\be \label{eq:Wkcmb}
W^{\kappa_{c}}=\frac{3}{2}\Omega_m H_0^2\frac{1+z}{H(z)}\chi(z)\times\frac{\chi(z_*)-\chi(z)}{\chi(z_*)} \, .
\ee
For the cosmic shear, we need to know the distribution of the source galaxies.
Beside the CMB lensing and cosmic shear angular power spectrum, we can plot the cross correlation of the CMB-cosmic shear as:
\be
C_l^{\kappa_{c}\kappa_{g}}=\int_0^\infty dz\frac{H(z)}{\chi^2(z)}W^{\kappa_{c}}(z)W^{\kappa_{g}}(z)P(k,z) \, .
\ee

In Fig. (\ref{Kernels}), we plot the kernels of the CMB lensing potential ($i=\phi$) introduced in Eq. (\ref{eq:Wphi}), CMB lensing convergence ($i=\kappa_{c}$) introduced in Eq. (\ref{eq:Wkcmb}) and the galaxy lensing convergence ($i=\kappa_{g}$) introduced in Eq.(\ref{eq:Wkg}) as a function of the redshift. An important point to indicate is that the kernels in the standard $\Lambda$CDM model are just functions of the redshift. The figure shows the effectiveness of each redshift bin on the overall signal. Also one must be cautious that the galaxy lensing kernel is very sensitive to the selection function $dn/dz$, which shows the number density of the observed galaxies in each redshift bin. This distribution is very well affected by the characteristics of the galaxy survey. For $W^{\kappa_g}$ we used the galaxy distribution from Hand et al. \cite{Hand:2013xua}. In the next section we will discuss the deviation from standard primordial conditions and study its effect on CMB and galaxy lensing.


\section{Deviation from the standard  primordial perturbations}
\label{Sec:Deviation}
As mentioned before the primordial fluctuations are nearly Gaussian, nearly scale invariant and  isotropic. In this section we propose a number of phenomenological models for primordial power spectrum,
which deviate from these standard assumptions.  Our goal is to study
the imprints of these deviations on  the CMB lensing.

\subsection{Hemispherical asymmetry: the general approach}

In Grishchuk and Zeldovich \cite{Grishchuk1978} there is a theoretical motivation to study the effect of the dipole asymmetry in primordial universe and later its implications on observable universe such as on CMB.  There were indications of dipole asymmetry in WMAP data \cite{Eriksen:2007pc}. The subsequent Planck observations also seem to support
the existence of dipole asymmetry \cite{Ade:2015hxq, Ade:2013nlj}, also see \cite{Aiola:2015rqa, Mukherjee:2015mma, Mukherjee:2015wra, Adhikari:2014mua} for follow up works on CMB analysis in the presence of dipole asymmetry.
This has caused significant interests on theoretical and observational
implications of hemispherical asymmetry. The simplest realization of hemispherical asymmetry is in the form of  dipole modulation
\be
\Delta T(\hat{n})= \overline { \Delta{T}}(\hat{n})\left[1+ A_d\,  {\hat{n}.\hat{p}} \right],
\ee
in which $\overline { \Delta{T}}(\hat{n})$ is the isotropic temperature fluctuation, $A_d$ is the amplitude of dipole,  $\hat{n}$ is the direction of the observation and $\hat p$ is the preferred direction in sky. The Planck team found
$\hat{p}=(\ell=227, b=-27)$ in galactic coordinate with the amplitude $A_d\simeq0.07$. In addition, the Planck data
indicates a non-trivial scale dependence for dipole asymmetry, i.e. $A_d = A_d(k)$, such that the amplitude of dipole
falls off rapidly for angular scale $l > 100$.

There is no convincing theoretical understanding of hemispherical asymmetry. One promising approach is the
idea of long mode modulations  \cite{aniso-longmode}. In this picture
a long mode $k_L$, which is much bigger than the Hubble radius, causes the asymmetry by modulating the background inflationary parameters such as the inflaton field or its velocity or by modulating the surface of end of inflation.
However, it is not easy for this idea to work in simple models of inflation such as in single field models. Indeed it is shown
in \cite{Namjoo:2013fka,  Abolhasani:2013vaa, Namjoo:2014nra} that the amplitude of dipole modulation is related to the amplitude of local-type non-Gaussianity $f_{NL}$. In single field models of inflation with small (actually zero $f_{NL}$) there is no chance to generate
 dipole asymmetry with large enough amplitude. Therefore, one has to look for beyond simple slow roll models of inflation employing ideas such as multiple  fields models like curvaton scenarios, iso-curvature perturbations, domain walls,
 non-vacuum inflationary initial conditions  etc. For a list of various theoretical works on these directions see  \cite{various}. If  the power asymmetry is not a statistical fluke
 it indicates towards a non-trivial early universe physics.

Now using the fact that ${\cal{P}}_{\cal{R}}\sim \langle \Delta T^2(\hat{n}) \rangle$ the primordial power spectrum  is obtained to be
\be
\label{dipole}
{\cal{P}}_{{\cal{R}}}=\mathcal {P_R}^{iso}(k)  \Big[1+2A_d \, {\hat{n}.\hat{p}} \Big]  \, .
\ee
In this section we would like to study the effect of this dipole modulation on CMB lensing.  However, before that we extend the above dipole modulation to more general modulation considering  azimuthally symmetric in which
\begin{align}
\label{PR-general}
 \mathcal {P_R}(k)= \mathcal {P_R}^{iso}(k)   \Big[1+ \sum_{l''m''} f_{l''m''}(k)Y_{l''m''}(\theta) \Big],
\end{align}
where $\theta$ is the angle between the patch in the sky and the preference direction and $f_{l''m''}$ is the coefficient which determines the amplitude of anisotropy and in general can be a function of wavenumber. The specific case of $(l'',m'')=(1,0)$ gives the dipole modulation.

In order to obtain the angular power spectrum of CMB lensing potential, we use Eqs. (\ref{120}) and (\ref{123}) :
\begin{eqnarray}
& \sum\limits_{lml'm'}\langle \phi^*_{lm}\phi_{l'm'}\rangle Y_{l'm'}(\widehat{n'})Y^*_{lm}(\widehat{n})  =16\pi \sum\limits_{lml'm'}\int_0^{\chi_{*}}  d\chi  \int_0^{\chi_{*}}d\chi'  (\dfrac{\chi_{*}-\chi}{\chi_{*} \chi})   (\dfrac{\chi_{*}-\chi'}{\chi_{*}\chi'}) \times    \int \dfrac{dk}{k} {\cal{ T}}(k;\eta){\cal{T}}(k;\eta')  \mathcal {P_R}^{iso}(k)
\nonumber\\&
~~~~~~~~~~~~~~~~~~~~~~~~~~~~\times  \Big[1+ \sum\limits_{l''m''} f_{l''m''}(k)Y_{l''m''}(\theta) \Big] j_l(k\chi)  j_{l'}(k\chi') Y^*_{lm}(\widehat{n}) Y_{l'm'}(\widehat{n'})\delta_{ll'} \delta_{mm'}   \label{148}
\end{eqnarray}
The anisotropic angular power spectrum $C_{lm ,l' m'}^{\phi\phi}$ is defined as below
\begin{align} \label{eq:clphi}
\langle \phi_{lm}\phi^*_{l'm'}\rangle=C_{lm ,l' m'}^{\phi\phi}
\end{align}
The isotropic part is the same as Eq.(\ref{130}). For the anisotropic contribution, we use an approximation to simplify the equation. Suppose we perform the statistics in a patch in the sky which has the angle $\theta$ relative to the preference direction. We assume  the patch is large enough  to produce large enough statistics and small enough  to be considered isotropic in the sky. This approximation is valid as long as the variation in asymmetry power spectrum across the sky is small.  Considering the anisotropic part of Eq. (\ref{148}) we obtain
\begin{eqnarray}
&\sum\limits_{lml'm'}\langle \phi^*_{lm},\phi_{l'm'}\rangle Y_{l'm'}(\widehat{n'})Y^*_{lm}(\widehat{n})  =16\pi \sum\limits_{lml'm'}\int_0^{\chi_{*}}  d\chi  \int_0^{\chi_{*}}d\chi'  (\dfrac{\chi_{*}-\chi}{\chi_{*} \chi})   (\dfrac{\chi_{*}-\chi'}{\chi_{*}\chi'})  \int \dfrac{dk}{k} {\cal{T}}(k;\eta){\cal{T}}(k;\eta')
\nonumber\\&
~~~~~~~~~~~~~~~~~~~~~~~~~~~~~~~~~~~~~~~~~~\times     \mathcal {P_R}^{iso}(k)  \sum_{l''m''} f_{l''m''}Y_{l''m''}(\theta) j_l(k\chi)  j_{l'}(k\chi') Y^*_{lm}(\widehat{n}) Y_{l'm'}(\widehat{n'})\delta_{ll'} \delta_{mm'} \label{154}
\end{eqnarray}
Using statistical isotropy in each patch $\langle \phi_{lm} \phi^*_{l'm'} \rangle \simeq \delta_{ll'} \delta_{mm'} C_l^{\phi\phi} (\theta)$.
 Now by expanding $C_l^{\phi\phi}$ in spherical harmonics space with respect to $\theta$
\begin{align}
C_l^{\phi\phi}(\theta) \simeq \sum\limits _{l''m''} C_{ll''m''}^{\phi\phi} Y_{l''m''} (\theta) \, ,
\end{align}
and rewriting  Eq. (\ref{154}) in terms of the new coefficient of expansion, we obtain

\begin{align}
C_{ll''m''}^{\phi\phi} (\theta)= 16\pi \int_0^{\chi_{*}}  d\chi  \int_0^{\chi_{*}}d\chi'  (\dfrac{\chi_{*}-\chi}{\chi_{*} \chi})   (\dfrac{\chi_{*}-\chi'}{\chi_{*}\chi'})
\times    \int \dfrac{dk}{k}  {\cal{T}}(k;\eta){\cal{T}}(k;\eta')  \mathcal {P_R}^{iso}(k)  f_{l''m''}   j_l(k\chi)  j_{l}(k\chi')  \, .
\label{Cllm}
\end{align}
For different patches the coefficients $ f_{l''m''}$ are different, so $ f_{l''m''}$ is a function of  $\theta$. As it is obvious from Eq. (\ref{Cllm}), the asymmetry in lensing power spectrum is induced from the asymmetric
nature of the primordial curvature perturbation power spectrum in  Eq. (\ref{PR-general}).

Having presented the general form of lensing power spectrum for arbitrary shape of asymmetry, in the next sub-section we study the special case of dipole modulation.


\subsection{Dipole asymmetry}

In the special case of dipole asymmetry in the primordial power spectrum we have
\begin{align}
 \mathcal {P_R}(k)= \mathcal {P_R}^{iso}(k) \Big[1+ A(k)  \widehat{\bf p}.\widehat {\bf n} \Big] =\mathcal {P_R}^{iso}(k) \Big[1+2 A(k) \sqrt{\frac{\pi}{3}} Y_{10}(\theta) \Big].
\end{align}
Note that we have allowed for the scale dependence of dipole amplitude $A(k)$.

Now considering the fact that the dipole modulation is a specific sample of general anisotropy in real space,  from
Eq. (\ref{Cllm})  we obtain
\begin{align}
C_{l10}^{\phi\phi} (\theta)= 16\pi \int_0^{\chi_{*}}  d\chi  \int_0^{\chi_{*}}d\chi'  (\dfrac{\chi_{*}-\chi}{\chi_{*} \chi})   (\dfrac{\chi_{*}-\chi'}{\chi_{*}\chi'})
\times    \int \dfrac{dk}{k}  {\cal{T}}(k;\eta){\cal{T}}(k;\eta')  \mathcal {P_R}^{iso}(k)  A(k)   j_l(k\chi)  j_{l}(k\chi') \, .
\end{align}
where $f_{10} $ is replaced by $A(k)$.

One important issue to consider is the scale dependence of dipole asymmetry.  As mentioned before
the CMB data shows that the dipole is effective only on large scales (small moments). In order to quantify this scale dependence we consider the amplitude of the dipole modulation $A(k)$ to have the following scaling form
\be \label{eq:adnd}
 A(k)= {{{A_d}}} \left(\dfrac{k}{k_d} \right)^{n_d-1},
\ee
where $A_d$ is a constant,  $k_d$ represents the scales at which the anisotropy is damped and $n_d$ is the spectral index of dipole power spectrum.
In this work we set the value of the pivot wavenumber of dipole modulation to $k_d = 0.015 h/Mpc$. This value is translated to the moment of $\ell \simeq 40$, where we assume is the mean value of moment which CMB dipole modulation dies off.


\begin{figure}
\advance\leftskip-1.5cm
\includegraphics[scale=1.2]{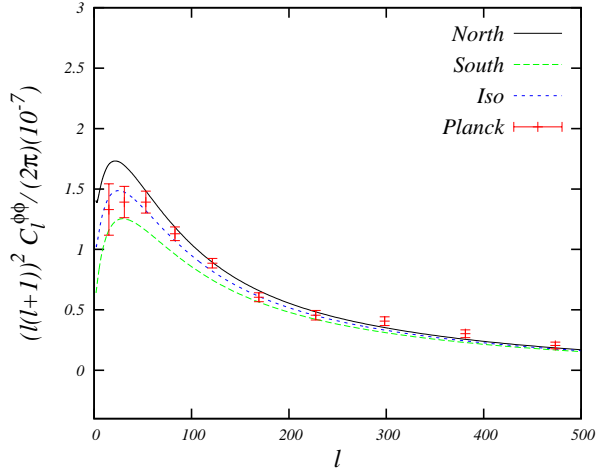}
\caption{ The angular power spectrum of CMB lensing potential for the isotropic case (blue dashed curve), for northern hemisphere  (black solid curve) and for southern hemisphere (green long dashed  curve) with  $A_d=0.07$,  $n_d=0.5$ and $k_d=0.015 h/Mpc$. The data points are from Planck lensing map reconstruction.}\label{fig:Clphi-di-data-norm}
\end{figure}


One can compare $C_{l}^{\phi\phi}(\theta)$ in different directions in order to probe the amplitude of dipole modulation. It is also possible to test the scale dependence of $ A(k)$. Taking into account the scale dependent anisotropy in CAMB code one can estimate the lensing potential power spectrum for different patches in the sky and check the consistency of model with observations. We consider the observable quantity introduced  below in order to check the prediction of anisotropic model on CMB lensing potential
\be
\label{Cl-NS}
C_l^{\phi\phi(N-S)}/\bar{C}_l=\dfrac{C_l^{North}-C_l^{South}}{\bar{C}_l} \,
\ee
where $C_l^{North}$ and $C_l^{South}$ are the powers in northern and southern hemispheres of the sky  defined with respect to the asymmetric  direction of dipole and $\bar{C}_l$ is the isotropic power spectrum. Note that
$C_l^{\phi\phi(N-S)}$ is an observable quantity. By comparing between the northern and southern hemispheres of CMB sky, one can obtain this quantity from observation.  In appendix \ref{app1} we investigate that how the $C_l^{\phi\phi(N-S)}/\bar{C}_l$ will depend on the parameters of dipole modulation, like dipole amplitude, the spectral index of dipole modulation and also the effect of the observing patch. \\

 On the other hand, theoretically, we can obtain  $C_l^{\phi\phi(N-S)}$  from the  CAMB code and  we can examine the consistency of the model.
In Fig. (\ref{fig:Clphi-di-data-norm}) we plot the full sky angular power spectrum of CMB lensing potential for the $\Lambda$CDM case and the cases with
the maximum power modulation  (northern hemisphere) and minimum power modulation (southern hemisphere)  with  $A_d=0.07$, $n_d=0.5$ and $k_d=0.015 h/Mpc$.
Fig. (\ref{fig:Clphi-di-data-norm}) shows that the maximum deviation from standard case is in low multipole ($l<100$), however the analysis of large patches to detect the dipole modulation is difficult due to limitation from cosmic variance.

Beside the CMB lensing potential, we can investigate the effect of the dipole modulation on the CMB lensing convergence as well. For this purpose, and similar to  $C_l^{\phi\phi(N-S)}$ defined in Eq. (\ref{Cl-NS}), we define  $C_l^{\kappa_c\kappa_c(N-S)}$ as the difference of the convergence angular power spectrum between the  northern and southern hemispheres. In appendix \ref{app1}, we discuss the dependence of $C_l^{\kappa_c\kappa_c(N-S)}$ on the spectral index of the dipole modulation.


\begin{figure}
\advance\leftskip-1.5cm
\includegraphics[scale=1.2]{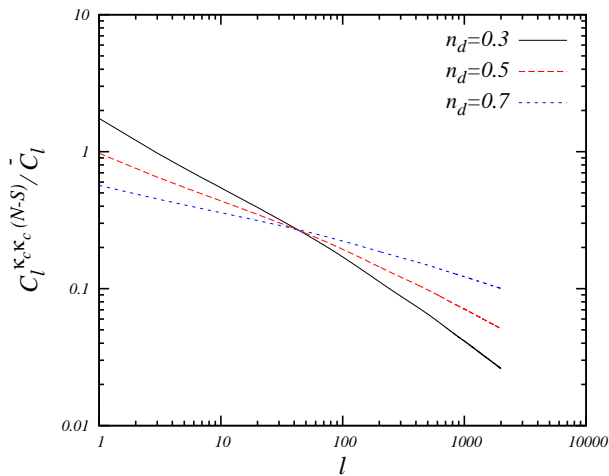}
\caption{ The fractional difference of  the  angular power of CMB lensing convergence in northern and southern hemispheres of CMB sky compared to the standard case $C_l^{\kappa_c\kappa_c(N-S)}/\bar C_l$: For $n_d=0.3$ (black solid curve), $n_d=0.5$ (red long dashed curve)  and $n_d=0.7$ (blue dashed curve) with $A_d=0.07$ and $k_d=0.015 h/Mpc$ for all curves.}\label{fig:kcmbkcmb-dipole-nd}
\end{figure}


 In Fig. (\ref{fig:kcmbkcmb-dipole-nd}), we present the effects of $n_d$ on $C_l^{\kappa_c\kappa_c(N-S)}$ , it is obvious that the behavior of the CMB convergence power spectrum is very similar to the lensing power, accordingly the effect of change in amplitude  and the dependence of the power to $\theta$  is very similar to Figs. (\ref{ClpsiAmp}) and (\ref{Clphidegree}). However the CMB convergence can be used as a complementary observation for lensing potential power to probe modulation and its scale dependence.



\begin{figure}
\advance\leftskip-1.5cm
\includegraphics[scale=1.2]{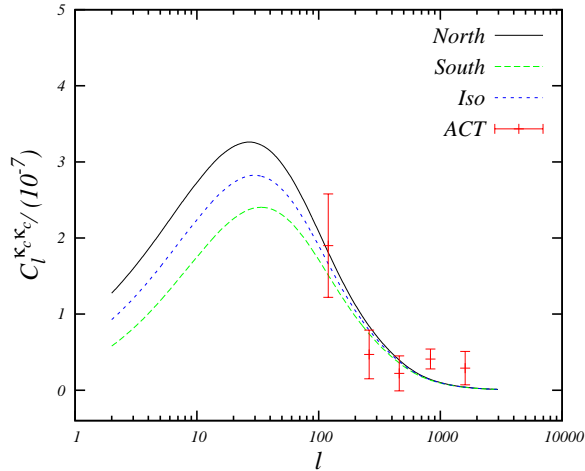}
\caption{  The angular power spectrum of CMB lensing convergence versus moments in logarithmic scale: the isotropic case (blue dashed line), northern hemisphere (black solid line), southern hemisphere  (green long dashed  line) with  $A_d=0.07$,  $n_d=0.5$ and $k_d=0.015 h/Mpc$. The data points are from ACT project \cite{Hand:2013xua}.}\label{fig:kcmbkcmb-dipole-data-log}
\end{figure}


In Fig. (\ref{fig:kcmbkcmb-dipole-data-log}) we plot $C_l^{\kappa_c\kappa_c}$ for  the northern and southern hemispheres with $n_d=0.5$, $k_d=0.0 1/Mpc$ and $A_d=0.07$. We compare the results with the data from Atacama Cosmology Telescope (ACT) \cite{Hand:2013xua}. The important point is that the  most significant deviation occurs in low multipoles while the convergence data from ACT can span the high multipoles. This is because  the telescope has a field of view of $22'\times 26 '$ which scan the CMB in three frequencies in a seasonal way. However,  the small field of view of the telescope will help to map the CMB lensing in different patches with different angles with respect to the dipole direction. Accordingly, our proposal for detecting the signal of dipole modulation is to study the CMB lensing convergence in each patch of observational sky separately without combining all the data.


\begin{figure}
\advance\leftskip-1.5cm
\includegraphics[scale=1.2]{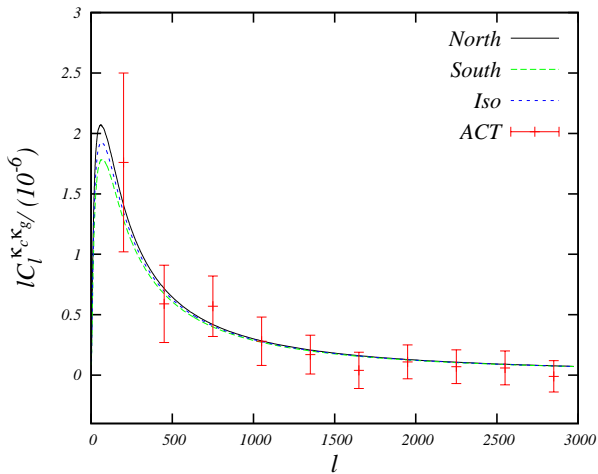}
 \caption{The cross angular power spectrum of CMB lensing convergence and galaxy lensing convergence: the isotropic case (blue dashed line), the northern hemisphere (black solid line) and  the southern hemisphere (green long dashed  line) with  $A_d=0.07$,  $n_d=0.5$ and $k_d=0.015 h/Mpc$. The data points are from ACT project \cite{Hand:2013xua}.}\label{fig:kcmbkgal-dipole-data-norm}
\end{figure}


In the search for dipole asymmetry in late time lensing data,  we can also use the cosmic shear observation and
use the cross-correlation of the convergence of galaxy lensing with CMB lensing. The cross-correlation of the CMB lensing and galaxy lensing permits us to probe the signal of modulation for larger angular moments as well. For this purpose, and similar to the case of lensing potential $C_l^{\phi\phi(N-S)}$,  we define $C_l^{\kappa_c\kappa_g(N-S)}$ as a measure of the difference of the cross angular power spectrum of CMB lensing convergence and galaxy lensing convergence between two hemispheres. The physics of cross-correlation and its relation to the dipole asymmetry parameters is studied in App. \ref{app1}.

In Fig. (\ref{fig:kcmbkgal-dipole-data-norm})  we compare the signal of cross-correlation of CMB lensing and galaxy lensing $C_l^{k_ck_g}$. The curve descriptions are the same as in Fig. (\ref{fig:kcmbkcmb-dipole-data-log}).
The data points are from the joint analysis of the ACT data for CMB lensing and CFHTLens data for galaxy lensing \cite{Hand:2013xua}. An important point to indicate is that the  cross angular power spectrum of CMB lensing convergence and galaxy lensing convergence depends on the Kernel of the galaxies, accordingly each galaxy sample can change the moment dependence of the signal due to its specific Kernel. This is important because future LSS surveys will provide us with new catalogs of galaxy lensing in different redshifts and accordingly different Kernels.

\begin{figure}
\advance\leftskip-1.5cm
\includegraphics[scale=1.2]{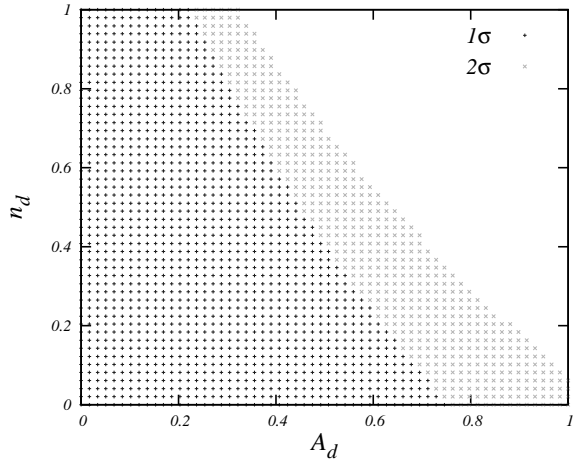}
 \caption{The $1\sigma$ and $2\sigma$ confidence levels of the amplitude of dipole $A_d$ and the spectral index $n_d$ are plotted  using the ACT convergence data and ACT-CFTHLens cross correlation of CMB convergence, galaxy convergence. }\label{fig:rev-counter}
\end{figure}

As a final word in this subsection we use the ACT convergence data and cross correlation of CMB lensing and cosmic shear data \cite{Hand:2013xua} to put constrains on the parameters of dipole modulation. We use only the ACT data for CMB lensing as this project maps a patch of sky and not the whole CMB sky. The Planck sky takes the average of all angles with respect to the dipole direction.
The ACT data is taken from a strip - survey of  galactic coordinate of right ascension $(01h ,06h )$ and declination $(-54^{\circ} ,-50^{\circ} )$ \cite{Das:2010ga}. We convert this area in galactic coordinate and by approximation we assume that the whole patch (center of the patch as a representative)  has a cosine angle of $\cos\theta_{\hat{p}.\hat{n}} \sim 0.71 $  with dipole direction $(l,b)$ . The lensing amplitude is fixed with the data for each sample and then we put the constrain in $A_d$ and $n_d$ defined in Eq.(\ref{eq:adnd}). In Fig. (\ref{fig:rev-counter}) we plot the $1\sigma$ and $2\sigma$ confidence levels.


The Figure shows that the data are consistent with null detection of dipole asymmetry. However we should keep in mind that the data points are very sparse with large error-bars. Future CMB lensing - cosmic shear data can tighten the constraint on the dipole parameters. Also  Fig.(\ref{fig:rev-counter}) shows that recent data put an upper threshold on the amplitude of dipole modulation with a specific spectral $n_d$.


\subsection{Deviation from Scale Invariance}


One way to check the deviation from the standard initial conditions for primordial perturbations is to study the  behavior of primordial power in different wavenumbers. Scale invariance is a key property of primordial fluctuations, which is observed from the scales of CMB down to the scale of group of galaxies at  $k\simeq 1 h/ Mpc$. However, there is no direct test of scale invariance
or its violation on smaller scales.   In this sub-section we study the effect of a strong violation of scale invariance
 in primordial power spectrum  on small scales  $k \sim 1 h/Mpc$ which is now out of the access of large scale structure surveys.

As a toy model, we consider a model of primordial power spectrum with localized feature at a  specific wavenumber $k_f$ (the subscript  $f$ stands for feature) in the following form
\begin{equation} \label{eq:powerbump}
 \mathcal {P_R}(k)= \mathcal {\overline{P}_R}(k)  \Big[1+{{A}}_f  \delta(k-k_f) \Big]  \, ,
\end{equation}
where $\mathcal {\overline{P}_R}(k)$ is the standard power spectrum of $\Lambda$CDM model with isotropic, scale invariance and Gaussian initial conditions.  In these parameterizations ${{A}}_f$ measures  the amplitude of the local feature
at the wavenumber $k_f$.
Our goal is to see how  this local feature affect the observation of the CMB-lensing. For a relevant study concerning
the effects of local feature on the abundance of large scale structures see  \cite{Baghram:2014nha}.

Here  let us pause to provide the theoretical motivation for generating  local feature and its naturalness in inflationary dynamics. In its simplest realization, inflation is driven by a single field slowly rolling over a flat potential. In this simple picture, the primordial power spectrum is very nearly scale-invariant in all observable scales. However, there is no fundamental reason that inflationary dynamics should follow this simple picture. Indeed, it is quite reasonable that there are many fields during inflation which can affect the dynamics of inflaton field sometime during its evolution. In particular, if some of the fields trigger instabilities  in field space, then this instability may imprint itself as a local violation of slow roll conditions on inflaton dynamics. This local violation of slow roll conditions can be captured effectively as a local glitch  in power spectrum as presented in Eq. (\ref{eq:powerbump}). For a realization of these kinds of inflationary scenarios  see \cite{Abolhasani:2012px}. Of course the local feature represented by Eq. (\ref{eq:powerbump}) is
oversimplified and a real primordial feature may have a more non-trivial shape than this highly localized shape.
Having that said, we use Eq. (\ref{eq:powerbump}) as a toy model which can help us to perform the analysis analytically.
As we shall see, in our numerical analysis and plots  we extend this artificial feature into a more realistic feature in which the bump has a Gaussian width.

Inserting the primordial power spectrum given in Eq. (\ref{eq:powerbump}) in the angular power spectrum of lensing potential defined in Eq.(\ref{eq:clpsi}), we can calculate the modified power spectrum of CMB-lensing potential. Keeping in mind that the power spectrum  does not depend on the direction of wavenumber $\hat{\bf k}$  we can separate the integral over the comoving part and the wavenumber part as follows
\begin{equation}
C_l^{\phi\phi} =16\pi \int \dfrac{dk}{k}\mathcal {\overline{P}_R}(k)  \left(1+{A}_f \delta(k-k_f) \right) \left\lbrack\int_0^{\chi_{*}}  d\chi (\dfrac{\chi_{*}-\chi}{\chi_{*} \chi})    T(k;\eta)  j_l(k\chi)  \right \rbrack ^2 \, .
\end{equation}
Now we can calculate the modification induced from the local feature  $\Delta C^{\phi\phi(f)}_l$. This modification is a function of ${A_f}$, $k_f$ and also the standard model parameters defined as below
\be
\Delta C^{\phi\phi(f)}_l({A_f},k_f)\equiv C_l^{\phi\phi}-\bar{C}_l^{\phi\phi}= \dfrac{16\pi  A_f}{k_f}  \mathcal {{\overline{P}}_R} (k_f) \left\lbrack\int_0^{\chi_{*}}  d\chi (\dfrac{\chi_{*}-\chi}{\chi_{*} \chi})    T(k_f;\eta)  j_l(k_f\chi)   \right\rbrack ^2 \, ,
\ee
where $\bar{C}_l^{\phi\phi}$ is the angular power spectrum of lensing potential in the standard case, i.e. in the absence of local feature.


\begin{figure}
\advance\leftskip-1.5cm
\includegraphics[scale=1.2]{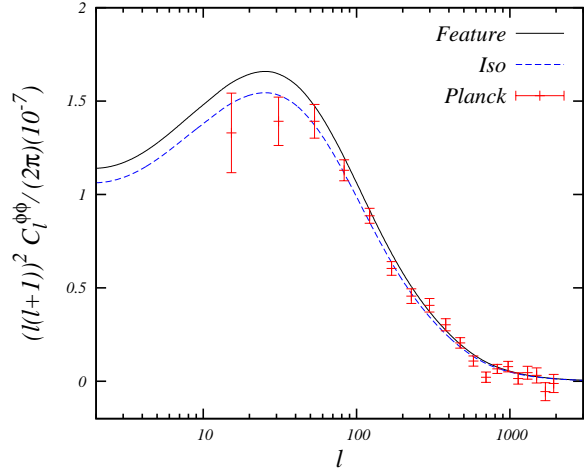}
\caption{The CMB lensing potential is plotted  in logarithmic scales for standard case of $\Lambda$CDM (blue dashed line) and for a Gaussian feature with $A_f=0.3h/Mpc$, $k_f=1 h/Mpc$ and  $\sigma_f=1 h/Mpc$. The data point is taken from lensing reconstruction of Planck data.}\label{Clphi-data-log}
\end{figure}


\begin{figure}
\advance\leftskip-1.5cm
\includegraphics[scale=1.2]{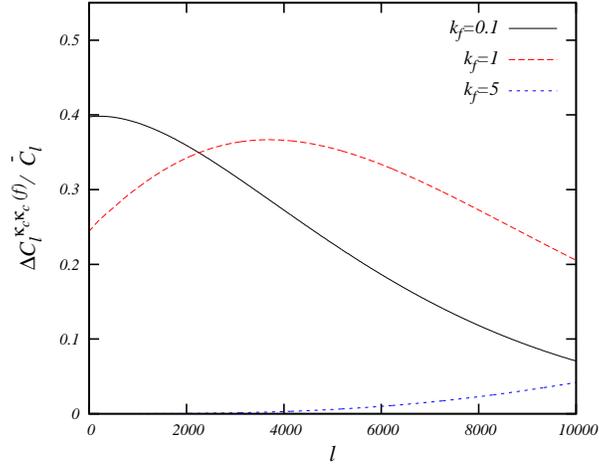}
\caption{The  the fractional change  in angular power spectrum of lensing convergence,
$\Delta C_l^{\kappa_c\kappa_c(f)}/\bar C_l$ for $k_f=0.1 h/Mpc$ (solid black curve), $k_f=1 h/Mpc$ (red long dashed  curve) and $k_f=5 h/Mpc$ (blue dashed curve) with $A_f=1 h/Mpc$ and $\sigma_f=1 h/Mpc$ in all curves. }\label{Clkckc-f-kf}
\end{figure}



\begin{figure}
\advance\leftskip-1.5cm
\includegraphics[scale=1.2]{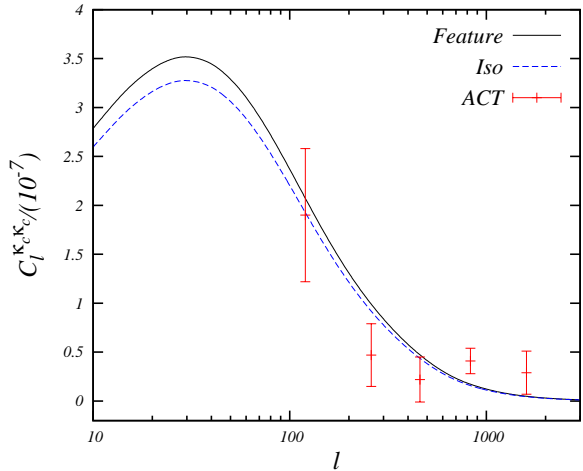}
\caption{The CMB lensing convergence  in logarithmic scale for standard case of $\Lambda$CDM (blue dashed curve) and  a Gaussian feature with  $A_f=0.3$, $k_f=1 h/Mpc$ and $\sigma_f=1 h/Mpc$ (black solid curve). The data point is taken from ACT data. }\label{Clkckc-data-log}
\end{figure}


As mentioned before, the local feature presented in Eq. (\ref{eq:powerbump}) is over-simplified mainly to allow
us to perform the above integrals analytically. However, in order to study more realistic cases, we can change the Dirac delta-function in Eq. (\ref{eq:powerbump}) by a Gaussian function defined as
\be\label{Eq:gauss}
\Delta{\cal{P}}_{{\cal{R}}}=\frac{{{A}}_f { {\mathcal {\overline{P}_R}(k) }}}{\sqrt{2\pi}\sigma_f}e^{-(k-k_f)^2/2\sigma^2_f} \, ,
\ee
where $\sigma_f$ is the width of the Gaussian function of the feature and $k_f$ is the wavenumber where the feature is centered. We use the above realistic feature in our numerical plots.
In Fig. (\ref{Clphi-data-log}) we have plotted the angular power spectrum of CMB lensing potential in log scale. The data points are from the Planck collaboration. The prediction of
$\Lambda$CDM model with standard initial conditions and the  Gaussian feature with  $\sigma_f=1 h/Mpc$, $A_f=0.3h/Mpc$ and $k_f=1 h/Mpc$ are jointly plotted. Generally a feature with positive amplitude increases the amplitude of power. The dependence of the CMB lensing potential to the parameters of power spectrum with a Gaussian feature is discussed and plotted in detail in Appendix (\ref{app2}).
Now we can also investigate the effect of the feature on the angular power spectrum of lensing convergence.
In Fig. (\ref{Clkckc-f-kf})  the fractional change  in angular power spectrum of lensing convergence,
$\Delta C_l^{\kappa_c\kappa_c(f)}/\bar C_l$, is plotted for different values of  characteristic wavenumber $k_f$. The dependence of  this fractional change in convergence power spectrum  to the amplitude and the width of the feature is the same as in the CMB lensing potential. The important point to mention here is that the general behavior of the lensing convergence signal is the same as lensing potential.

In Fig. (\ref{Clkckc-data-log}) we plot the CMB lensing convergence power spectrum
$\Delta C_l^{\kappa_c\kappa_c(f)}$ for the Gaussian feature with  $k_f=1 h/Mpc$, $A_f=0.3 h/Mpc$ and $\sigma_f= 1 h/Mpc$. The convergence data points are from ACT collaboration\cite{Hand:2013xua}.

As we discussed in the dipole modulation case study, the cross correlation of the CMB lensing and galaxy lensing improves the observational opportunity to test the deviations from the isotropic and scale invariant power spectrum
for smaller scales (higher moments). Accordingly we study the effect of feature on the cross correlation power
$\Delta C_l^{\kappa_c\kappa_g(f)}$.


\begin{figure}
\advance\leftskip-1.5cm
\includegraphics[scale=1.2]{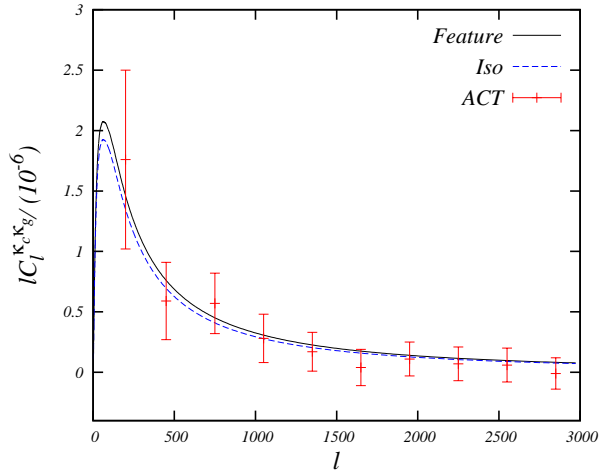}
\caption{The cross correlation of the CMB lensing convergence and galaxy lensing convergence  for standard case of $\Lambda$CDM (blue dashed curve) and for the case of Gaussian feature with $A_f=0.3 h/Mpc$,  $k_f=1 h/Mpc$  and $\sigma_f=1 h/Mpc$ (black solid curve). The data point is taken from ACT collaboration \cite{Hand:2013xua}}\label{Clkckg-data-norm}
\end{figure}


In Fig. (\ref{Clkckg-data-norm}), we plot  the angular power
spectrum of the cross correlation of CMB lensing and galaxy lensing for both the standard case and the feature model with  $A_f=0.3 h/Mpc$,  $k_f=1 h/Mpc$ and $\sigma_f=1 h/Mpc$. The data points are taken from the ACT data.

 A very important point to indicate is that the convergence data has a higher resolution and it spans the moments up to $ l \simeq 2500$. Accordingly, the lensing convergence is a more prominent observation to test the deviation from scale invariance by the excess in amplitude on smaller scales.


\section{Summary and Conclusions}
\label{Sec-Conclusion}

Cosmological observations strongly support the standard model of cosmology with the  initial conditions which are nearly isotropic, Gaussian, adiabatic and  scale invariant. However it is crucial to test the properties of initial conditions with different observations and on different scales. In this work we have proposed  that the CMB lensing can be a prominent new observational tool to investigate the physics of initial conditions. The CMB lensing can be a unique way to test these deviations. This is because the  map of
last scattering photons is distorted by  the structures between the observer and CMB while  these late time structures  carry rich  information about the physics of  early universe encoded in  their statistical properties.
In addition,  we also suggested that the cross-correlation of CMB and galaxy lensing can also be used as a probe of initial conditions. In the case of the cross-correlation, each galaxy sample probes a different range of wavenumbers due to its corresponding kernel which is related to the distribution of the galaxies.

There are indications of hemispherical asymmetry in CMB observations.   We have proposed that both the CMB lensing potential and CMB lensing convergence can be used to probe the amplitude of dipole modulation. An important feature is that the amplitude of the CMB lensing potential is sensitive to both the amplitude of dipole and also to  the orientation  with respect to the direction of dipole modulation. The   future CMB observations are planned to scan  the last scattering surface with high resolution in small patches. Accordingly they can be used to address the predictions of this study. In addition the CMB lensing can probe the scale dependence of the dipole modulation. The Planck data suggests a non-trivial scale-dependence for dipole amplitude in which the amplitude of dipole falls off rapidly for $l > 60$.  We showed that the CMB lensing potential (convergence)  can be used to constrain  the  scale dependence of dipole modulation. In this work we considered the simple scaling relation  $\Delta{\cal{P}}={{A_f}} \left(k/k_d\right)^{(n_d-1)}$ for dipole amplitude and have plotted the signal of CMB lensing as a function of the spectral index of dipole modulation. We have used the CMB lensing convergence autocorrelation  and the cosmic shear convergence cross correlation with CMB to constrain the amplitude of $A_f$ and the spectral index of dipole modulation.

Another example of deviation from the standard primordial initial conditions which can be tested via  CMB lensing is the deviation from scale invariance. The observational data from CMB and LSS show that up to scales of groups of galaxies the primordial power spectrum is  scale invariant. However,  there is no direct evidence that this scale invariance continues on smaller scales.  In the second part of this work we have studied the question of deviation from  scale invariance in the form of local feature in primordial power spectrum.   In order to model this feature we have assumed a Gaussian-type bump in the primordial power on small scales $k> 1 h/Mpc$.
We studied the effect of the amplitude of the bump, the location of feature in momentum space $k_f$
and the effect of the variance of Gaussian bump on the CMB lensing potential and convergence.  We also studied the effect of feature on the cross correlation of CMB lensing convergence and galaxy lensing.
We showed that this cross correlation has the ability to probe the signals in higher moments (small scales) in examining
the scale dependence of primordial power spectrum.

There are other questions which one can study in this direction. For example, one can study the question of statistical anisotropy in primordial power spectrum. This can arise for example in models of inflation containing background vector fields and gauge fields, for a review see \cite{Emami:2015qjl}. Because of the background vector field, the rotational symmetry is broken. The primordial power spectrum has the form
\be
\label{SI}
 \mathcal {P_R}(k)= \mathcal {\bar{P}_R}  \left[1+ g_* \left( \widehat {\bf v}\cdot  \widehat {\bf k}\right)^2 \right] \, ,
\ee
in which ${\bf \widehat v}$ represents the preferred direction in the sky and $g_*$ measures the amplitude of quadrupole anisotropy. The  date from Planck observations yield  the constrain $| g_*| \lesssim 10^{-2}$ \cite{Ade:2013uln, Kim:2013gka}.   Note that the anisotropy in Eq. (\ref{SI}) has a different nature
than the dipole asymmetry in Eq. (\ref{dipole}). Statistical anisotropy given in Eq. (\ref{SI}) is defined in Fourier space and measures anisotropy
for each point in the sky while dipole asymmetry given in Eq. (\ref{dipole}) holds in real space and only distinguishes two opposite hemispheres and not the individual points. It will be interesting to use the CMB lensing potential and CMB lensing convergence to probe the statistical anisotropy of the primordial perturbations.

\acknowledgments
We would like to thank Anthony Lewis, Anthony Challinor and Yashar Akrami  for insightful comments.  We also thank the anonymous referee for his/her detailed and insightful comments and suggestions which helped  to improve the contents and the presentation of the manuscript.

\appendix
\section{Phenomenological predictions for Dipole asymmetry in CMB sky }
\label{app1}

 In this Appendix we investigate the parameter space of the dipole asymmetry effect on the lensing potential correlation function and also on the cross correlation of the CMB convergence and cosmic shear lensing. The idea is to use the  observable $C_l^{\phi\phi(N-S)}/\bar{C}_l=\dfrac{C_l^{North}-C_l^{South}}{\bar{C}_l} \, $ which is introduced in Sec. (\ref{Sec:Deviation}) as a probe to pin down the physics of dipole asymmetry.



\begin{figure}[!htb]
\minipage[t]{0.31\linewidth}
\advance\leftskip-1.5cm
\includegraphics[width=6.3cm, height=5.0cm]{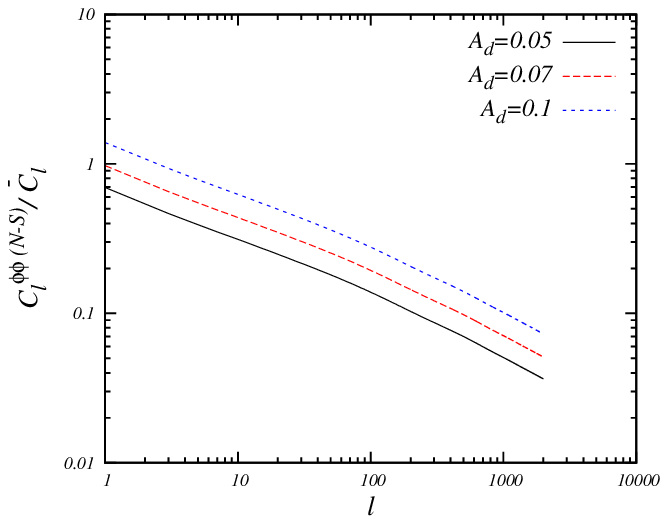}
\caption{The fractional difference in  angular power of CMB lensing potential in northern and southern hemispheres of CMB sky compared to the standard case, $C_l^{\phi\phi(N-S)}/\bar{C}_l$  as defined in Eq. (\ref{Cl-NS}) is plotted versus moments. The plots are for amplitudes $A_d=0.05$ (black solid line), $A_d=0.07$ (red long dashed line) and $A_d=0.1$ (blue dashed line). In all case the spectral index of dipole modulation is $n_d=0.5$ and the pivot wavenumber of dipole modulation is set to $k_d=0.015 h/Mpc$  }\label{ClpsiAmp}
\endminipage\hfill
\minipage[t]{0.31\linewidth}
\advance\leftskip-1.5cm
\includegraphics[width=6.3cm, height=5.0cm]{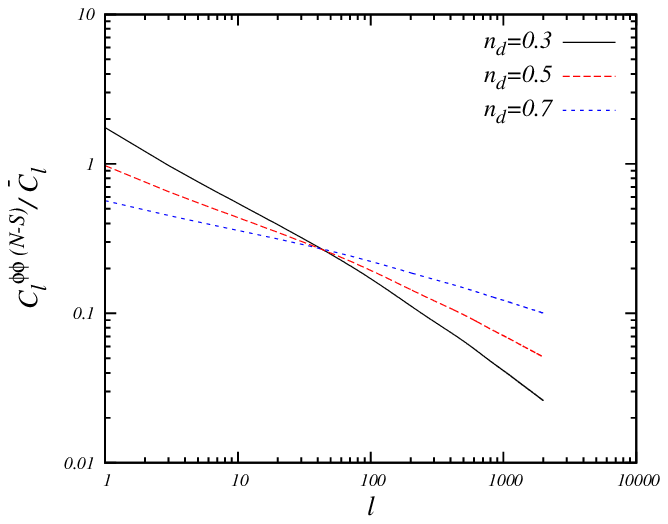}
\caption{The plot of $C_l^{\phi\phi(N-S)}/\bar{C}_l$ for $n_d=0.3$ (black solid curve), $n_d=0.5$ (red long dashed curve) and $n_d=0.7$ (blue dashed curve) with $A_d=0.07$ and $k_d=0.015 h/Mpc$ for all curves.
}\label{Clphiindex}
\endminipage\hfill
\minipage[t]{0.31\linewidth}
\advance\leftskip-1.5cm
\includegraphics[width=6.3cm, height=5.0cm]{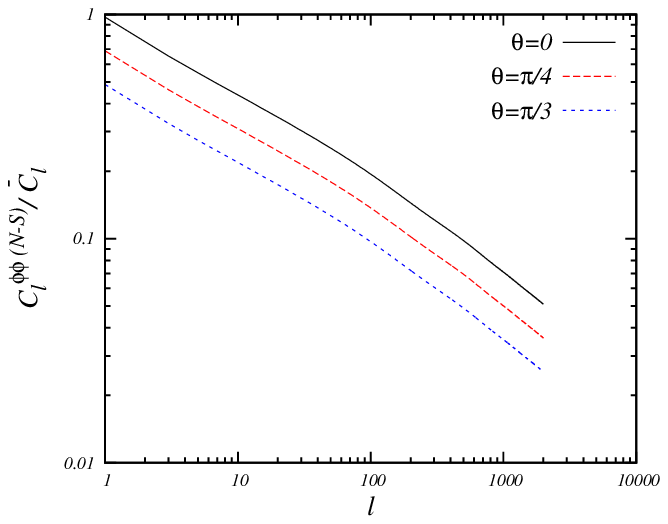}
\caption{The plot of $C_l^{\phi\phi(N-S)}$ for different values of $\theta$: $\theta=0.0$ (black solid curve),  $\theta= \pi/4$  (red long dashed curve)  and  $\theta=\pi/3$   (blue dashed curve)  with  $A_d=0.07$ and $k_d=0.015 h/Mpc$ for all curves.} \label{Clphidegree}
\endminipage \hfill
\end{figure}

In Fig. (\ref{ClpsiAmp})  $C_l^{\phi\phi(N-S)}/\bar{C}_l$, as defined in Eq. (\ref{Cl-NS}), is plotted
versus the moment $l$ for different values of dipole amplitude:  $A_d=0.05$ (black solid curve), $A_d=0.07$ (red long dashed curve) and  $A_d=0.1$ (blue dashed curve). In all cases the spectral index of dipole modulation is set to $n_d=0.5$ and the pivot wavenumber of dipole modulation is set to $k_d=0.015 h/Mpc$. The figure shows that probing the CMB lensing in two opposite hemispheres can constrain the amplitude of the dipole modulation.
 In Fig. (\ref{Clphiindex}), $C_l^{\phi\phi(N-S)}/\bar{C}_l$ is plotted for different values of $n_d$: $n_d=0.3$ (black solid curve),
$n_d=0.5$ (red long dashed curve),  $n_d=0.7$ (blue dashed curve). In all cases  $A_d=0.07$ and  $k_d=0.015 h/Mpc$.
The Figure shows that a smaller spectral index introduces more negative slope for the power and accordingly the difference in  dipole modulation in low and high momenta are more enhanced. The fixed point where all powers have the same magnitude represents the pivot wavenumber.
As described before the pivot number is chosen to have the pivot moment of $\ell_p\sim 40$.
Accordingly we assert that the CMB lensing is a promising tool to constrain the scale dependence of the dipole modulation. Measuring the high and low moments of lensing potential can be used as  criteria to constrain the spectral index of the modulation.

In Fig. (\ref{Clphidegree}), $C_l^{\phi\phi(N-S)}/\bar{C}_l$ is plotted for different values of $\theta$:  $\theta=0.0$ (black solid curve),  $\theta= \pi/4$ (red long dashed curve) and  $\theta=\pi/3$ (blue dashed curve) in which in all plots $A_d=0.07$ and  $k_d=0.015  h/Mpc$. It is obvious that the maximum difference can be obtained in direction of dipole.


\begin{figure}[!htb]
\minipage[t]{0.31\textwidth}
\advance\leftskip-1.4cm
\includegraphics[width=6.5cm, height=5.0cm]{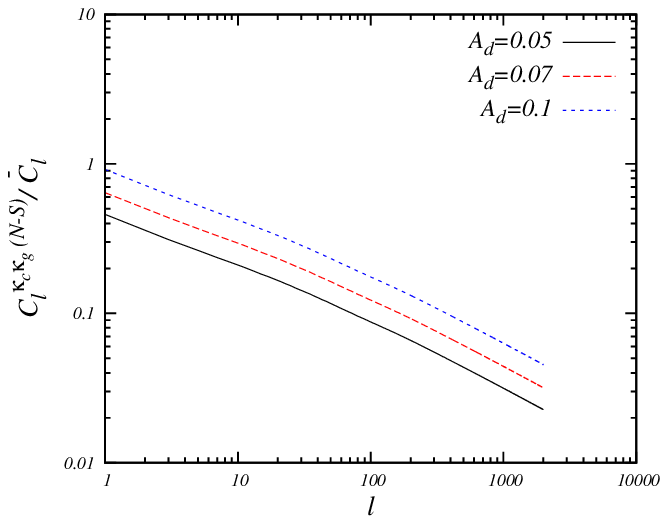}
\caption{The plot of fractional difference for CMB lensing convergence  in northern and southern hemispheres of CMB sky compared to the standard case $C_l^{\kappa_c\kappa_g(N-S)}$:  $A_d=0.05$ (black solid curve), $A_d=0.07$ (red long dashed curve), $A_d=0.1$ (blue dashed curve) with $n_d=0.5$ and $k_d=0.015 h/Mpc$ for all curves.}
\label{fig:kcmbkgal-dipole-amp}
\endminipage\hfill
\minipage[t]{0.31\textwidth}
\advance\leftskip-1.4cm
\includegraphics[width=6.5cm, height=5.0cm, center]{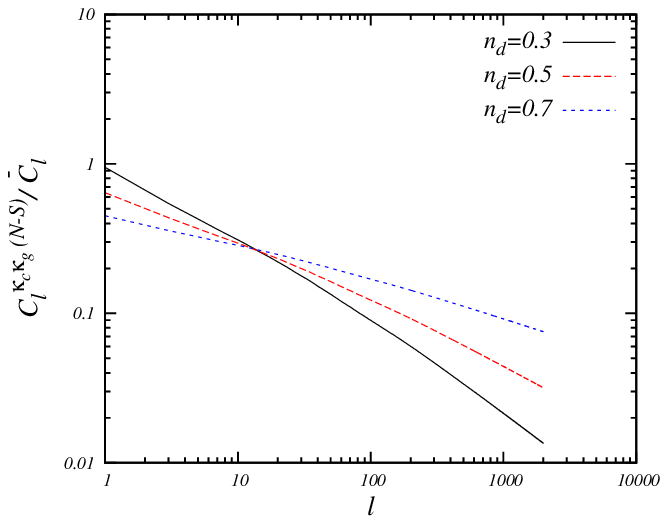}
\caption{The plot of $C_l^{\kappa_c\kappa_g(N-S)}$ for $n_d=0.3$ (black solid curve), $n_d=0.5$ (red long dashed curve)  and $n_d=0.7$ (blue dashed curve) with $A_d=0.07$ and $k_d=0.015 h/Mpc$ for all curves. }
\label{fig:kcmbkgal-dipole-nd}
\endminipage\hfill
\minipage[t]{0.31\textwidth}
\advance\leftskip-0.7cm
\includegraphics[width=6.5cm, height=5.0cm, right]{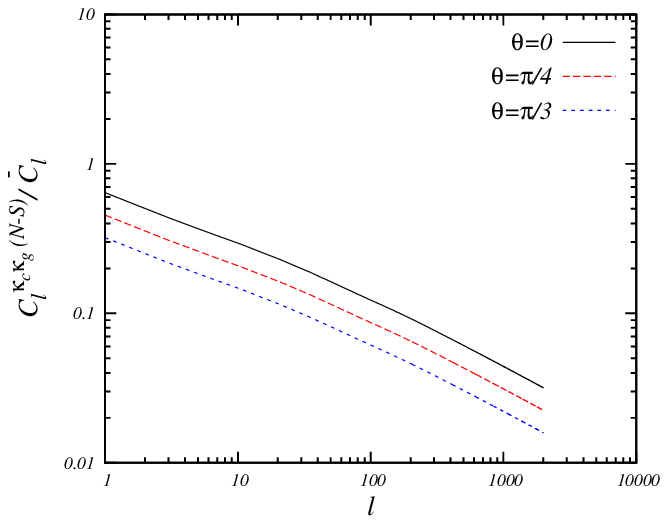}
\caption{The  The plot of $C_l^{\kappa_c\kappa_g(N-S)}$  for different values of $\theta$: $\theta=0.0$ (black solid curve),  $\theta= \pi/4$  (red long dashed curve)  and  $\theta=\pi/3$   (blue dashed curve)  with  $A_d=0.07$ and $k_d=0.015  h/Mpc$ for all curves.}\label{fig:kcmbkgal-dipole-degree}
\endminipage\hfill
\end{figure}

As we discussed extensively in Sec. (\ref{Sec:Deviation}), another promising tool to study the deviation from standard primordial perturbation is the cross correlation of CMB convergence with cosmic shear data. A crucial point to indicate here is that each cross correlation signal is affected by the distribution of the galaxies, where it can probe the dipole asymmetry in different moments.
In Figs. (\ref{fig:kcmbkgal-dipole-amp}),  (\ref{fig:kcmbkgal-dipole-nd}) and  (\ref{fig:kcmbkgal-dipole-degree})
we study respectively the effect of $A_d$, $n_d$ and $\theta$ on $C_l^{\kappa_c\kappa_g(N-S)}$. The curve descriptions respectively  are the same as in Figs. (\ref{ClpsiAmp}),   (\ref{Clphiindex}) and (\ref{Clphidegree}).

\section{Phenomenological predictions for feature in CMB sky }
\label{app2}

In this Appendix we probe the parameter space of a Gaussian feature in primordial power spectrum with CMB lensing and CMB convergence - galaxy convergence cross correlation data. The idea is the same as raised in App. (\ref{app1})  where we plot the difference in powers of the lensing and convergence
for the cases with the feature compared with the standard one.


\begin{figure}[!htb]
\minipage[t]{0.31\textwidth}
\advance\leftskip-1.7cm
\includegraphics[width=6.5cm, height=5.0cm, left]{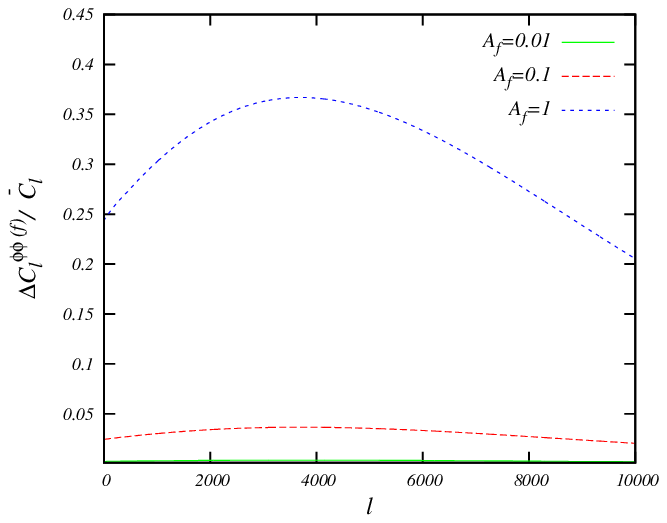}
\caption{The fractional change in CMB lensing potential due to feature $\Delta C_l^{\phi\phi (f)}/\bar{C}_l$:
  $A_f=0.01 h/Mpc$ (solid green curve),  $A_f=0.1 h/Mpc$ (red long dashed curve) and   $A_f=1h/Mpc$ (blue dashed curve)
  with $k_f=1 h/Mpc$ and $\sigma_f=1 h/Mpc$ for all curves.}\label{Clphi-f-amp}
\endminipage\hfill
\minipage[t]{0.31\textwidth}
\advance\leftskip-1.5cm
\includegraphics[width=6.5cm, height=5.0cm, center]{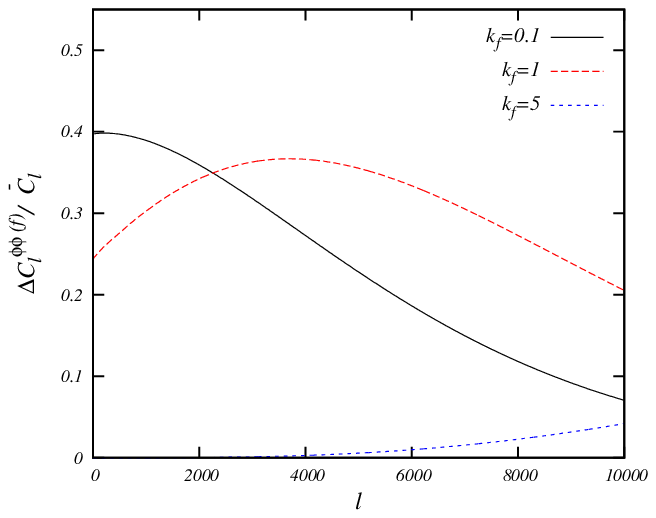}
\caption{The plot of $\Delta C_l^{\phi\phi (f)}/\bar{C}_l$:  $k_f=0.1 h/Mpc$ (solid black curve), $k_f=1 h/Mpc$ (red long dashed curve) and  $k_f=5 h/Mpc$ ( blue dashed curve) with  $A_f=1 h/Mpc$ and $\sigma_f=1 h/Mpc$ for all curves.}\label{Clphi-f-kf}
\endminipage\hfill
\minipage[t]{0.31\textwidth}
\advance\leftskip-1.0cm
\includegraphics[width=6.5cm, height=5.0cm, right]{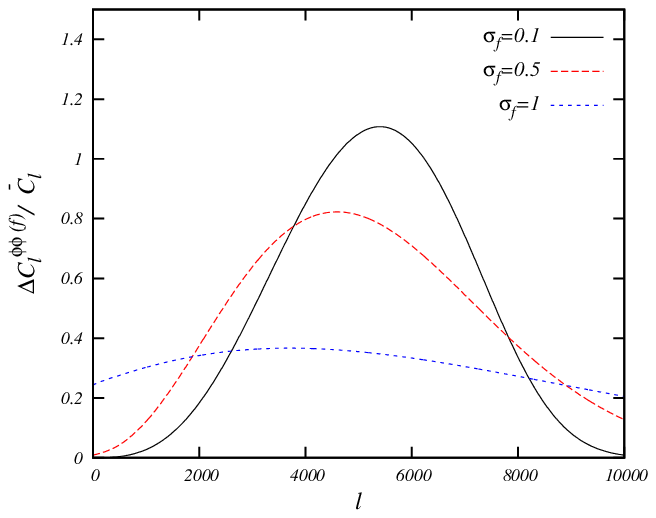}
\caption{The plot of $\Delta C_l^{\phi\phi (f)}/\bar{C}_l$:  $\sigma_f=0.1 h/Mpc$  (solid black curve),  $\sigma_f=0.5 h/Mpc$  (red long dashed curve) and  $\sigma_f=1 h/Mpc$  ( blue dashed curve) with $A_f=1 h/Mpc$ and $k_f=1 h/Mpc$
  for all curves.  }\label{Clphi-f-sigmaf}
\endminipage\hfill
\end{figure}

In Fig. (\ref{Clphi-f-amp}), the fractional change in lensing potential due to feature, $\Delta C_l^{\phi\phi (f)}/\bar{C}_l$, is plotted.  The feature has   Gaussian shape as in Eq. (\ref{Eq:gauss}) with $k_f= 1 h/Mpc$ and $\sigma_f= 1 h/Mpc$ held fixed. Plots are  for different values of  the amplitude,  $A_f=1 h/Mpc$ (blue dotted line), $A_f=0.1 h/Mpc$ (red long dashed line) and $A_f=0.01 h/Mpc$ (green solid line). Fig. (\ref{Clphi-f-amp}) shows that for features with large amplitudes the  CMB lensing can probe the deviation from the standard case.

In Fig. (\ref{Clphi-f-kf}) we have plotted $\Delta C_l^{\phi\phi (f)}/\bar{C}_l$  for the feature with $A_f=1 h/Mpc$ and $\sigma_f= 1 h/Mpc$ for different wavenumbers. An important fact is that by going to smaller scales (larger wavenumbers), the signal of deviation from scale invariance
appears in higher moments $l$. Accordingly, if we want to probe the deviation of the initial power spectrum from scale invariance in smaller and smaller scales, we have to reconstruct the CMB lensing maps in higher moments with corresponding resolution. The highest moment of CMB lensing potential reconstruction is now in the moment of $l\simeq 1500$.

In Fig. (\ref{Clphi-f-sigmaf}), we have plotted the same quantity as in previous figures for different value of $\sigma_f$. The smaller values of variance lead to  features in the angular power spectrum which have
higher amplitudes and are more localized.



\begin{figure}[!htb]
\minipage[t]{0.32\textwidth}
\advance\leftskip-1.5cm
\includegraphics[width=6.5cm, height=5.0cm, left]{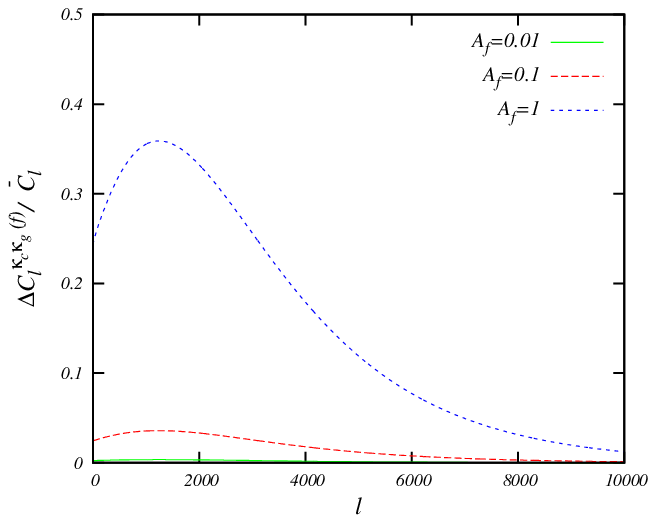}
\caption{The plot of $\Delta C_l^{\kappa_c\kappa_g(f)}/\bar C_l$ for  $A_f=0.01 h/Mpc$ ( solid green curve),   $A_f=0.1 h/Mpc$ (  red long dashed curve) and  $A_f=1 h/Mpc$ ( blue dashed curve) with  $k_f=1 h/Mpc$ and $\sigma_f=1 h/Mpc$ in all curves.}\label{Clkckg-f-amp}
\endminipage\hfill
\minipage[t]{0.32\textwidth}
\advance\leftskip-1.5cm
\includegraphics[width=6.5cm, height=5.0cm, center]{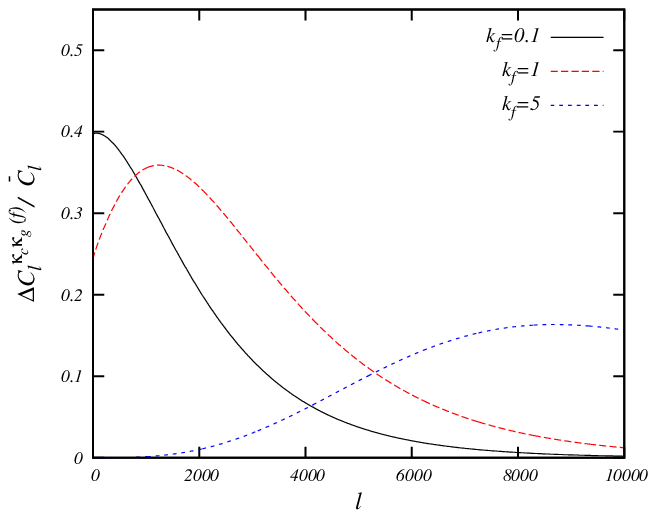}
\caption{The plot of $\Delta C_l^{\kappa_c\kappa_g(f)}/\bar C_l$ for   $k_f=0.1 h/Mpc$ ( solid black curve),
$k_f=1 h/Mpc$ ( red long dashed curve) and  $k_f=5 h/Mpc$ ( blue dashed curve) with $A_f=1 h/Mpc$ and $\sigma_f=1h/Mpc$ for all curves.}\label{Clkckg-f-kf}
\endminipage\hfill
\minipage[t]{0.32\textwidth}%
\advance\leftskip-1.0cm
\includegraphics[width=6.5cm, height=5.0cm, right]{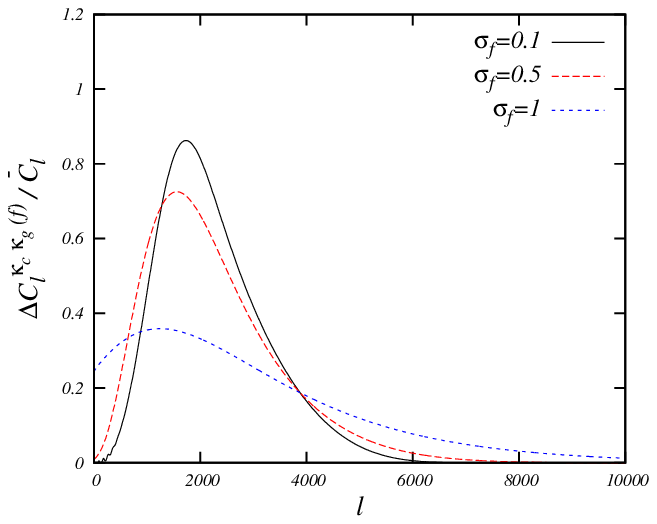}
\caption{The plot of $\Delta C_l^{\kappa_c\kappa_g(f)}/\bar C_l$ for    $\sigma_f=0.1 h/Mpc$ ( solid black  curve),
$\sigma_f=0.5 h/Mpc$ (  red long dashed curve) and  $\sigma_f=1 h/Mpc$  (blue dashed curve) with $A_f=1 h/Mpc$
and $k_f=1 h/Mpc$ for all curves.}\label{Clkckg-f-sigmaf}
\endminipage\hfill
\end{figure}

As mentioned we can also use the CMB convergence - galaxy convergence correlation data which in this case the signal is sourced by the distribution of galaxies as well.
In  Fig. (\ref{Clkckg-f-amp}) we plot $\Delta C_l^{\kappa_c\kappa_g(f)}/\bar C_l$
for different values of amplitude with  $k_f=1 h/Mpc$ and $\sigma_f=1 h/Mpc$. It is worth to mention that the peak  is shifted to smaller moments. In Fig. (\ref{Clkckg-f-kf}) and Fig. (\ref{Clkckg-f-sigmaf}) we study respectively the effects of change in $k_f$ and  $\sigma_f$ on $\Delta C_l^{\kappa_c\kappa_g(f)}/\bar C_l$.

\end{document}